\documentclass{article}
\pdfoutput=1
\usepackage[T1]{fontenc} % add special characters (e.g., umlaute)
\usepackage[utf8]{inputenc} % set utf-8 as default input encoding
\usepackage{ismir,amsmath,cite,url}
\usepackage{graphicx}
\usepackage{color}

\usepackage{xcolor,amsfonts}
\usepackage{booktabs}
\usepackage[caption=false]{subfig}

\title{Downbeat Tracking with Tempo-Invariant\\Convolutional Neural Networks}

\threeauthors
  {Bruno Di Giorgi} {Apple Inc. \\ {\tt bdigiorgi@apple.com}}
  {Matthias Mauch} {Apple Inc. \\ {\tt mmauch@apple.com}}
  {Mark Levy} {Apple Inc. \\ {\tt mark\_levy@apple.com}}

\begin{document}

\maketitle
\begin{abstract}
The human ability to track musical downbeats is robust to changes in tempo, and it extends to tempi never previously encountered.  We propose a deterministic time-warping operation that enables this skill in a convolutional neural network (CNN) by allowing the network to learn rhythmic patterns independently of tempo.  Unlike conventional deep learning approaches, which learn rhythmic patterns at the tempi present in the training dataset, the patterns learned in our model are tempo-invariant, leading to better tempo generalisation and more efficient usage of the network capacity.

We test the generalisation property on a synthetic dataset created by rendering the Groove MIDI Dataset using FluidSynth, split into a training set containing the original performances and a test set containing tempo-scaled versions rendered with different SoundFonts (test-time augmentation).
The proposed model generalises nearly perfectly to unseen tempi (F-measure of 0.89 on both training and test sets), whereas a comparable conventional CNN achieves similar accuracy only for the training set (0.89) and drops to 0.54 on the test set.
The generalisation advantage of the proposed model extends to real music, as shown by results on the GTZAN and Ballroom datasets.
\end{abstract}

\section{Introduction}  % Intro and literature
\label{sec:intro}

Human musicians easily identify the downbeat (the first beat of each bar) in a piece of music and will effortlessly adjust to a variety of tempi, even ones never before encountered.  This ability is the likely result of patterns and tempi being processed at distinct locations in the human brain \cite{thaut2014human}. 

We argue that factorising rhythm into tempo and tempo-invariant rhythmic patterns is desirable for a machine-learned downbeat detection system as much as it is for the human brain. First, factorised representations generally reduce the number of parameters that need to be learned. Second, having disentangled tempo from pattern we can transfer information learned for one tempo to all others, eliminating the need for training datasets to cover all combinations of tempo and pattern.

Identifying invariances to disentangle representations has proven useful in other domains \cite{higgins2018towards}: translation invariance was the main motivation behind CNNs \cite{lecun1995convolutional} --- the identity of a face should not depend on its position in an image. Similarly, voices retain many of their characteristics as pitch and level change, which can be exploited to predict pitch \cite{bittner2017deep} and vocal activity \cite{schluter2018zero}. Crucially, methods exploiting such invariances don't only generalise better than non-invariant models, they also perform better overall.

% Literature review with focus on how different algorithms deal with tempo invariance
Some beat and downbeat trackers first estimate tempo (or make use of a tempo oracle) and use the pre-calculated tempo information in the final tracking step \cite{davies2005beat,ellis2007beat,klapuri2005analysis,degara2011reliability,giorgi2016multipath,papadopoulos2010joint,krebs2016downbeat,davies2006spectral,durand2015downbeat,durand2016feature}. Doing so disentangles tempo and tempo-independent representations at the cost of propagating errors from the tempo estimation step to the final result.
It is therefore desirable to estimate tempo and phase simultaneously \cite{goto2001audio,dixon2001automatic,eck2007beat,goto2011songle,krebs2015inferring}, which however leads to a much larger parameter space. Factorising this space to make it amenable for machine learning is the core aim of this paper.

In recent years, many beat and downbeat tracking methods changed their front-end audio processing from hand-engineered onset detection functions towards beat-activation signals generated by neural networks \cite{bock2011enhanced,bock2016joint,korzeniowski2014probabilistic}. Deep learning architectures such as convolutional and recurrent neural networks are trained to directly classify the beat and downbeat frames, and therefore the resulting signal is usually cleaner. 

By extending the receptive field to several seconds, such architectures are able to identify rhythmic patterns at longer time scales, a prerequisite for predicting the downbeat. But conventional CNN implementations learn rhythmic patterns separately for each tempo, which introduces two problems. First, since datasets are biased towards mid-tempo songs, it introduces a tempo-bias that no post-processing stage can correct. Second, it stores similar rhythms redundantly, once for every relevant tempo, i.e.\ it makes inefficient use of network capacity.  Our proposed approach resolves these issues by learning rhythmic patterns that apply to all tempi.

% Proposed solution
The two technical contributions are as follows:
\begin{enumerate}
\item the introduction of a scale-invariant convolutional layer that learns temporal patterns irrespective of their scale.
\item the application of the scale-invariant convolutional layer to CNN-based downbeat tracking to explicitly learn tempo-invariant rhythmic patterns.
\end{enumerate}

Similar approaches to achieve scale-invariant CNNs, have been developed in the field of computer vision \cite{xu2014scale,kanazawa2014locally}, while no previous application exists for musical signal analysis, to the best of our knowledge. 

We demonstrate that the proposed method generalises better over unseen tempi and requires lower capacity with respect to a standard CNN-based downbeat tracker. The method also achieves good results against academic test sets.

\section{Model}
\label{sec:model}

The proposed downbeat tracking model has two components: a neural network to estimate the joint probability of downbeat presence and tempo for each time frame, using tempo-invariant convolution, and a hidden Markov model (HMM) to infer a globally optimal sequence of downbeat locations from the probability estimate. 

We discuss the proposed scale-invariant convolution in Sec.~\ref{sec:scaleinv} and its tempo-invariant application in Sec.~\ref{sec:tempoinv}. The entire neural network is described in Sec.~\ref{sec:network} and the post-processing HMM in Sec.~\ref{sec:post}.

\subsection{Scale-invariant convolutional layer}
\label{sec:scaleinv}

In order to achieve scale invariance we generalise the conventional convolutional neural network layer.

\subsubsection{Single-channel}
We explain this first in terms of a one-dimensional input tensor $x\in \mathbb{R}^N$ and only one kernel $h\in \mathbb{R}^{N^\ast}$, and later generalise the explanation to multiple channels in Sec.~\ref{sec:multichannel}. 
Conventional convolutional layers convolve $x$ with $h$ to obtain the output tensor $y\in\mathbb{R}^{N-N^\ast+1}$
\begin{align}
y = x \ast h,
\end{align}
\noindent where $\ast$ refers to the discrete convolution operation. Here, the kernel $h$ is updated directly during back-propagation, and there is no concept of scale. Any two patterns that are identical in all but scale (e.g.\ one is a ``stretched'' version of the other) cannot be represented by the same kernel.

\begin{figure}[t]
\centering
\includegraphics[width=1.0\columnwidth]{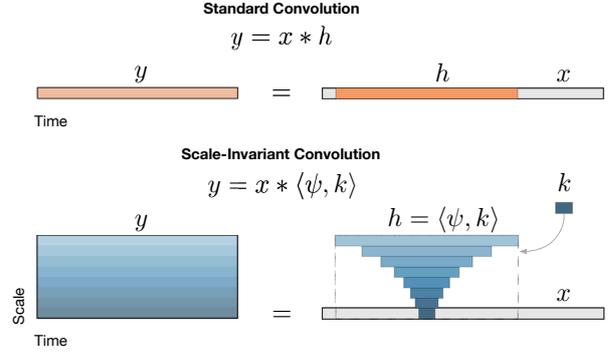}
\caption{
    The figure shows a representation of the standard and scale-invariant convolution operations with input/output channel dimensions removed for simplicity. In order to achieve scale invariance, we parametrise the kernel as the dot product of two tensors $\psi$ and $k$, where $\psi$ is a deterministic scaling tensor and $k$ is the trained part that will learn scale-invariant patterns. The resulting kernel $h$ contains multiple scaled versions of $k$.
}
\label{fig:tempoinv_a}
\end{figure}

To address this shortcoming, we factorise the kernel representation into scale and pattern by parametrising the kernel as the dot product $h_{j}=\langle\psi_j, k\rangle$ between a fixed scaling tensor $\psi_j\in\mathbb{R}^{N^\ast\times M}$ and a scale-invariant pattern $k\in \mathbb{R}^M$. Only the pattern is updated during network training, and the scaling tensor, corresponding to $S$ scaling matrices, is pre-calculated (Sec.~\ref{sec:scaling}). The operation adds an explicit scale dimension to the convolution output
\begin{align}
y_j = x \ast h_j = x\ast\langle\psi_j, k\rangle.
\label{eq:scaleinv_onechannel}
\end{align}
The convolution kernel is thus factorised into a constant scaling tensor $\psi$ and trainable weights $k$ that learn a scale-invariant pattern. A representation of a scale-invariant convolution is shown in Figure~\ref{fig:tempoinv_a}.

\begin{table}
\begin{tabular}{lcc}
\toprule
&\multicolumn{2}{c}{\textbf{layer input}} \\
\textbf {variable} & \textbf{single-channel} & \textbf{multi-channel}\\
\midrule
\# frames&\multicolumn{2}{c}{$N$}\\
\# pattern frames&\multicolumn{2}{c}{$M$}\\
\# scales&\multicolumn{2}{c}{$S$}\\
\# input channels & $1$ & $C_x$\\
\# kernels & $1$ & $H$\\
%\midrule
signal $x$ & $\mathbb{R}^N$ & $\mathbb{R}^{N\times C_x}$\\
patterns $k$&$\mathbb{R}^M$& $\mathbb{R}^{M\times C_x \times H}$\\
kernel $h$& $\mathbb{R}^{N^\ast\times S}$& $\mathbb{R}^{N^\ast\times C_x \times S \times H}$\\
output $y$&$\mathbb{R}^{(N-N^\ast+1)\times S}$&$\mathbb{R}^{(N-N^\ast+1) \times S \times H}$ \\
scaling tensor $\psi$ &\multicolumn{2}{c}{$\mathbb{R}^{N^\ast\times M \times S}$}\\

%\midrule
scale indices & \multicolumn{2}{c}{$j=0,\ldots,S-1$} \\
\bottomrule
\end{tabular}
\caption{Variables and dimensions.}
\label{tab:variables}
\end{table}

\subsubsection{Multi-channel}
\label{sec:multichannel}

Usually the input to the convolutional layer has $C_x>1$ input channels and there are $H > 1$ kernels. The formulas in Section~\ref{sec:scaleinv} can easily be extended by the channel dimension, as illustrated in Table~\ref{tab:variables}.

\subsubsection{Scaling tensor}
\label{sec:scaling}

The scaling tensor $\psi$ contains $S$ scaling matrices from size $M$ to $s_j M$ where $s_j$ are the scale factors.

\begin{equation}
    \label{eq:psi}
    \psi_{n,m,j} = \int_{\tilde{s}} \int_{\tilde{n}} \delta(\tilde{n}-\tilde{s} m)\kappa_n(n-\tilde{n})\kappa_s(s_j-\tilde{s})d\tilde{n}d\tilde{s},
\end{equation}

\noindent where $\delta$ is the Dirac delta function and $\kappa_n$, $\kappa_s$ are defined as follows: 

\begin{align*}
\kappa_n(d) &= \sin(\pi d)/(\pi d)\\
\kappa_s(d) &= \alpha \cos^2(\alpha d \pi / 2) \mathcal{H}(1 - \alpha|d|),
\end{align*}

\noindent
where $\mathcal{H}$ is the Heaviside step function.
The inner integral can be interpreted as computing a resampling matrix for a given scale factor and the outer integral as smoothing along the scale dimension, with the parameter $\alpha$ of the function $\kappa_s$ controlling the amount of smoothing applied.
The size $N^*$ of the scaling tensor $\psi$ (and the resulting convolutional kernel $h$) is derived from the most stretched version of $k$:

\begin{equation}
N^* = \max_j s_j M.
\end{equation}

\subsubsection{Stacking scale-invariant layers}
\label{sec:stacking}

After the first scale-invariant layer, the tensor has an additional dimension representing scale. In order to add further scale invariant convolutional layers without losing scale invariance, subsequent operations are applied scale-wise:

\begin{equation}
    y_j = x_j * \langle \psi_j, k \rangle.
    \label{eq:scaleinv_multiscale}
\end{equation}

\noindent The only difference with Eq.\ \eqref{eq:scaleinv_onechannel} is that the input tensor $x$ of Eq.\ \eqref{eq:scaleinv_multiscale} already contains $S$ scales, hence the added subscript $j$.

\subsection{Tempo invariance}
\label{sec:tempoinv}

In the context of the downbeat tracking task, tempo behaves as a scale factor and the tempo-invariant patterns are rhythmic patterns. We construct the sequence of scale factors $s$ as

\begin{equation}
    \label{eq:scale_to_tempo}
    s_j = \frac{r \tau_j B}{M},\quad \tau_j = \tau_0 2^{\frac{j}{T}}
\end{equation}

\noindent where $\tau_j$ are the beat periods, $r$ is the frame rate of the input feature, $B$ is the number of beats spanned by the convolution kernel factor $k$, $\tau_0$ is the shortest beat period, and $T$ is the desired number of tempo samples per octave. The matrix $k$ has a simple interpretation as a set of rhythm fragments in musical time with $M$ samples spanning $B$ beats.

\begin{figure}[t]
\centering
\includegraphics[width=0.6\columnwidth]{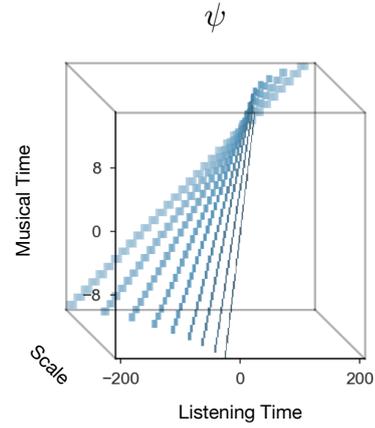}
\caption{
	The scaling tensor $\psi$ is a sparse 3-dimensional constant tensor. In the figure $\psi$ is represented as a cube where the 0 bins are rendered transparent. $\psi$ transforms the rhythm patterns contained in the kernel $k$ from musical time (e.g. 16th notes) to listening time (e.g. frames) over multiple scales.
}
\label{fig:tempoinv_b}
\end{figure}

To mimic our perception of tempo, the scale factors in Eq.\ \eqref{eq:scale_to_tempo} are log-spaced, therefore the integral in Eq.\ \eqref{eq:psi} becomes:

\begin{equation}
    \label{eq:psi_tempo}
    \psi_{n,m,j} = \int_{\tilde{j}} \int_{\tilde{n}} \delta(\tilde{n}-s_{\tilde{j}} m)\kappa_n(n-\tilde{n})\kappa_s(j-\tilde{j})d\tilde{n}d\tilde{j},
\end{equation}

\noindent 
where the parameter $\alpha$ of the function $\kappa_s$ has been set to $1$.
A representation of the scaling tensor used in the tempo-invariant convolution is shown in Figure~\ref{fig:tempoinv_b}.

\subsection{Network}
\label{sec:network}

\begin{figure}[t]
\centering
\includegraphics[width=0.8\columnwidth]{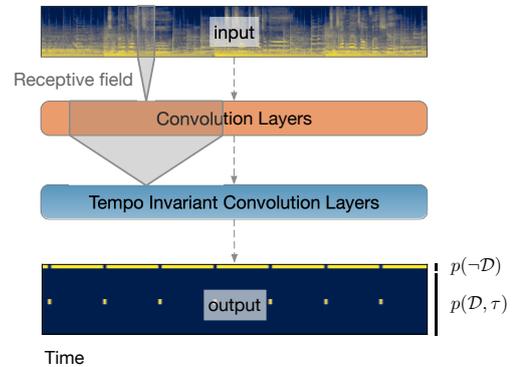}
\caption{
	A global view of the neural network. The first group of layers are regular convolutional layers and act as onset detectors. They have a small receptive field, in order to focus on acoustic features and avoid learning rhythmic patterns, which will be learned by the successive tempo-invariant layers. The output tensor represents joint probabilities of downbeat presence $\mathcal{D}$ and tempo $\tau$.
}
\label{fig:tempoinv_c}
\end{figure}

The tempo-invariant network (Fig.~\ref{fig:tempoinv_c}) is a fully convolutional deep neural network, where the layers are conceptually divided into two groups.
The first group of layers are regular one-dimensional convolutional layers and act as onset detectors. The receptive field is constrained in order to preserve the tempo-invariance property of the model: if even short rhythmic fragments are learned at a specific tempo, the invariance assumption would be violated. We limit the maximum size of the receptive field to 0.25 seconds, i.e.\ the period of a beat at 240 BPM.

The second group is a stack of tempo-invariant convolutional layers (as described in Sec.~\ref{sec:scaleinv}, \ref{sec:tempoinv}). The receptive field is measured in musical-time, with each layer spanning $B=4$ beats. 
The last layer outputs only one channel, producing a 2-dimensional (frame and tempo) output tensor.

The activations of the last layer represent the scores (logits) of having a downbeat at a specific tempo. An additional constant zero bin\footnote{We can keep this constant because the other output values will adapt automatically.} is concatenated to these activations for each frame to model the score of having no downbeat. After applying the softmax, the output $o$ represents the joint probability of the downbeat presence $\mathcal{D}$ at a specific tempo $\tau$
\begin{equation}
o_j = \left\{
        \begin{array}{ll}
            p(\mathcal{D}, \tau_j) \qquad & j = 0, \ldots , S-1 \\[1.0ex]
            p(\neg\mathcal{D}) \qquad & j = S
        \end{array}
    \right.
\end{equation}

\noindent The categorical cross-entropy loss is then applied frame-wise, with a weighting scheme that balances the loss contribution on downbeat versus non-downbeat frames.\footnote{The loss of non-downbeat frames is reduced to $1/3$.}

% Training
The target tensors are generated from the downbeat annotations by spreading the downbeat locations to the neighbouring time frames and tempi using a rectangular window ($0.1$ seconds wide) for time and a raised cosine window ($2/T$ octaves wide) for tempo. The network is trained with stochastic gradient descent using RMSprop, early stopping and learning rate reduction when the validation loss reaches a plateau.

\subsection{Post-processing}
\label{sec:post}

In order to transform the output activations of the network into a sequence of downbeat locations, we use a frame-wise HMM with the state-space \cite{krebs2015efficient}.

In its original form, this post-processing method uses a network activation that only encodes beat probability at each position. In the proposed tempo-invariant neural network the output activation models the joint probability of downbeat presence \emph{and tempo}, enabling a more explicit connection to the post-processing HMM, via a slightly modified observation model:

\begin{equation}
P(o_j | q) = \left\{
        \begin{array}{ll}
            c(\tau_j, \tau_q) o_j \qquad & q \in \mathcal{D},\quad j < S \\[1.0ex]
            o_S / (\sigma S) \qquad & q \in \neg\mathcal{D}
        \end{array}
    \right.
\end{equation}

\noindent where $q$ is the state variable having tempo $\tau_q$, $\mathcal{D}$ is the set of downbeat states, $c(\tau_j, \tau_q)$ is the interpolation coefficient from the tempi modeled by the network $\tau_j$ to the tempi modeled by the HMM $\tau_q$ and $\sigma$ approximates the proportion of non-downbeat and downbeat states ($|\neg\mathcal{D}| / |\mathcal{D}|$).

\section{Experiments}
\label{sec:experiments}

In this section we describe the two experiments conducted in order to test the tempo-invariance property of the proposed architecture with respect to a regular CNN. The first experiment, described in Sec.~\ref{ssec:ti_exp}, uses a synthetic dataset of drum MIDI recordings. The second experiment, outlined in Sec.~\ref{ssec:musicdata}, evaluates the potential of the proposed algorithm on real music.

\subsection{Tempo-invariance}
\label{ssec:ti_exp}

\begin{figure}[t]
\centering
\subfloat[accuracy with respect to relative tempo change]{
  \centering
  \centerline{\includegraphics[width=1.0\columnwidth]{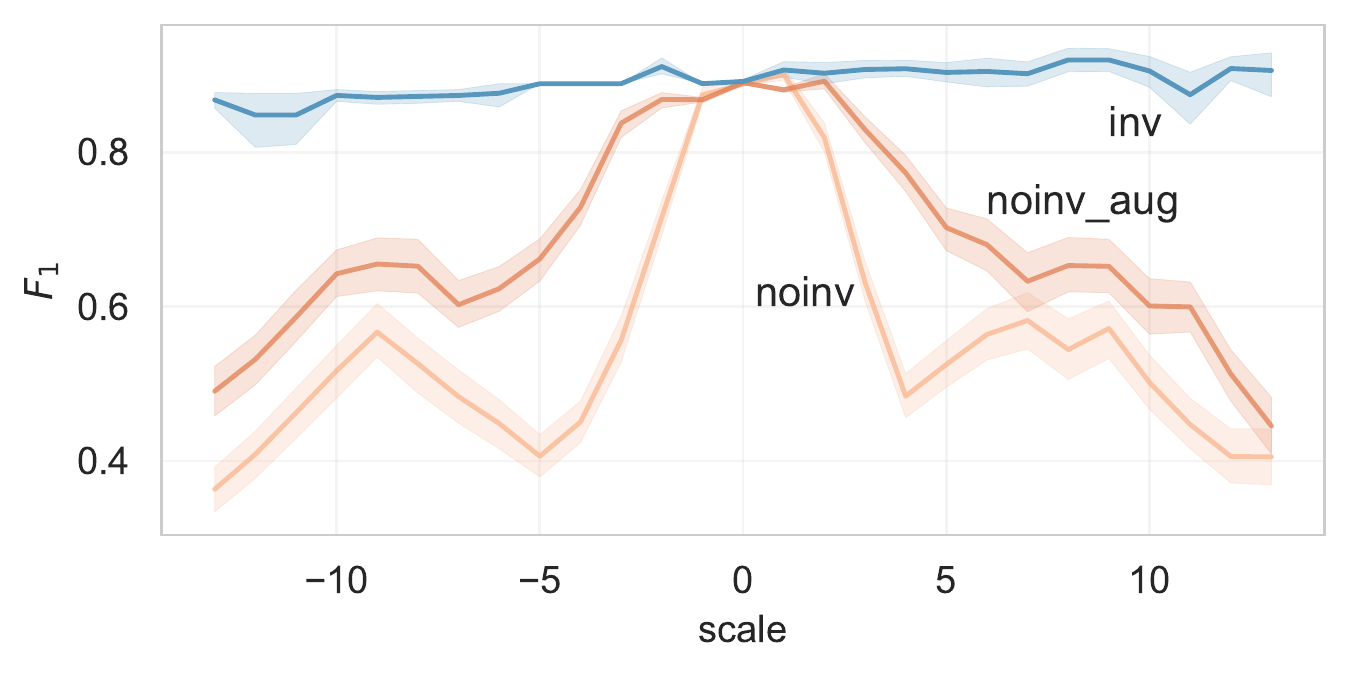}}
  \label{fig:tempoinvscale}
}
\newline
\subfloat[accuracy with respect to absolute tempo]{
  \centering
  \centerline{\includegraphics[width=1.0\columnwidth]{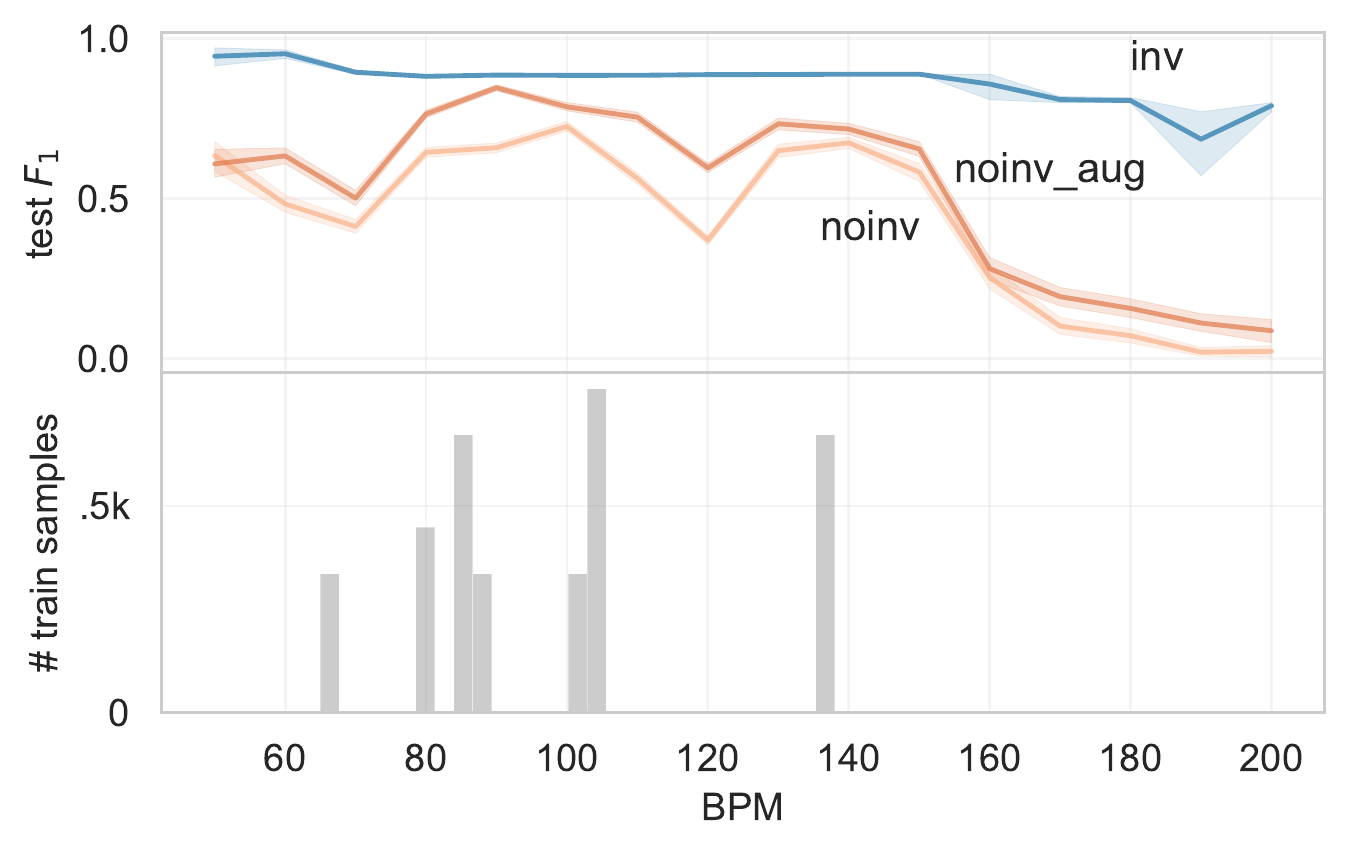}}
  \label{fig:tempoinvbpm}
}

\caption{
    Tempo invariance experiment using a dataset of $27$ time scaled versions of a set of drum patterns. The scale factors $\varepsilon_i=2^{i/26}$ range from $0.707$ to $1.414$.
    A tempo-invariant CNN (\texttt{inv}) and a standard CNN (\texttt{noinv}) are trained on the non scaled versions (scale=$0$) and tested on all others. A standard CNN trained on scales $[-1, 1]$ (\texttt{noinv\_aug}) simulates the effect of data augmentation. Figure~(a) shows that the invariant model is able to generalise on seen patterns at unseen tempi. Figure~(b) shows that the effect of the tempo-biased training set: for non-invariant models the benefit is localised, while the invariant model distributes the rhythmic information across the entire tempo spectrum.
}
\label{fig:tempoinvexperiment}
\end{figure}

We test the robustness of our model by training a regular CNN and a tempo-invariant CNN on a tempo-biased training dataset and evaluating on a tempo-unbiased test set. In order to control the tempo distribution of the dataset, we start with a set of MIDI drum patterns from the magenta-groove dataset \cite{groove2019}, randomly selecting $4$ bars from each of the $40$ \texttt{eval-sessions}, resulting in $160$ patterns. These rhythms were then synthesised at 27 scaled tempi, with scale factors $\varepsilon_i=2^{i/26}$ ($-13 \leq i \leq 13$) with respect to the original tempo of the recording. Each track starts with a short silence, the duration of which is randomly chosen within a bar length, after which the rhythm is repeated 4 times. Audio samples are rendered using FluidSynth\footnote{http://www.fluidsynth.org} with a set of $40$ combinations of SoundFonts\footnote{https://github.com/FluidSynth/fluidsynth/wiki/SoundFont} and instruments, resulting in $172800$ audio files.
The synthesised audio is pre-processed to obtain a log-amplitude mel-spectrogram with $64$ frequency bins and $r=50$ frames per second.

% dataset split
The tempo-biased training set contains the original tempi (scale factor: $\varepsilon_0=1$), while the tempo-unbiased test set contains all scaled versions. The two sets were rendered with different SoundFonts.

We compared a tempo-invariant architecture (\texttt{inv}) with a regular CNN (\texttt{noinv}).
The hyper-parameter configurations are shown in Table~\ref{tab:expconditions} and were selected maximising the accuracy on the validation set.

\begin{table}
\centering
\begin{tabular}{lcc}
\toprule
&\multicolumn{2}{c}{\textbf{architecture}} \\
\textbf{group} & \textbf{inv} & \textbf{noinv}\\
\midrule
1
&
\begin{tabular}{@{}c@{}}CNN \\ $3 \times 32$\end{tabular}
& 
\begin{tabular}{@{}c@{}}CNN \\ $3 \times 32$\end{tabular} \\[3ex]

2
&
\begin{tabular}{@{}c@{}}TI-CNN \\ 
$2 \times 16$ \\ 
$1 \times 1$ \\
\end{tabular}
& 
\begin{tabular}{@{}c@{}}dil-CNN \\
$3 \times 64$ \\
$1 \times 1$ \\
\end{tabular} \\
\midrule
\textbf{\#params} & 60k & 80k \\

\bottomrule
\end{tabular}
\caption{Architectures used in the experiment. Groups of layers are expressed as (number of layers $\times$ output channels). All layers in group 1 have kernel size equal to 3 frames. dil-CNN is a stack of dilated convolution layers with kernel size equal to 7 frames and exponentially increasing dilation factors: $2, 4, 8, 16$. The specific hyper-parameters of the tempo-invariant network TI-CNN are configured as follows: $T=8, \tau_0=0.25, S=25, M=64, B=4$. ReLU non-linearities are used on both architectures.}
\label{tab:expconditions}
\end{table}

% results of the exeperiment
The results of the experiment are shown in Fig.~\ref{fig:tempoinvexperiment} in terms of $F_1$ score, using the standard distance threshold of $70$ ms on both sides of the annotated downbeats \cite{davies2009evaluation}.
Despite the tempo bias of the training set, the accuracy of the proposed tempo-invariant architecture is approximately constant across the tempo spectrum. Conversely, the non-invariant CNN performs better on the tempi that are present in the training and validation set. Specifically, Fig.~\ref{fig:tempoinvscale} shows that the two architectures perform equally well on the training set containing the rhythms at their original tempo (scale equal to 0 in the figure), while the accuracy of the non-invariant network drops for the scaled versions.
A different view of the same results on Fig.~\ref{fig:tempoinvbpm} highlights how the test set accuracy depends on the scaled tempo. The accuracy of the regular CNN peaks around the tempi that are present in the training set, showing that the contribution of the training samples is localised in tempo. The proposed architecture performs better (even at the tempi that are present in the training set) because it efficiently distributes the benefit of all training samples over all tempi.

% augmentation experiment
In order to simulate the effect of data augmentation on the non-invariant model, we also trained an instance of the non-invariant model (\texttt{noinv\_aug}) including two scaled versions ($\varepsilon_i$ with $|i| \leq 1$) in the training set. As shown in the figure, data-augmentation improves generalisation, but has similar tempo dependency effects.

\subsection{Music data}
\label{ssec:musicdata}

\begin{figure}[t]
    \centering
    \includegraphics[width=1.0\columnwidth]{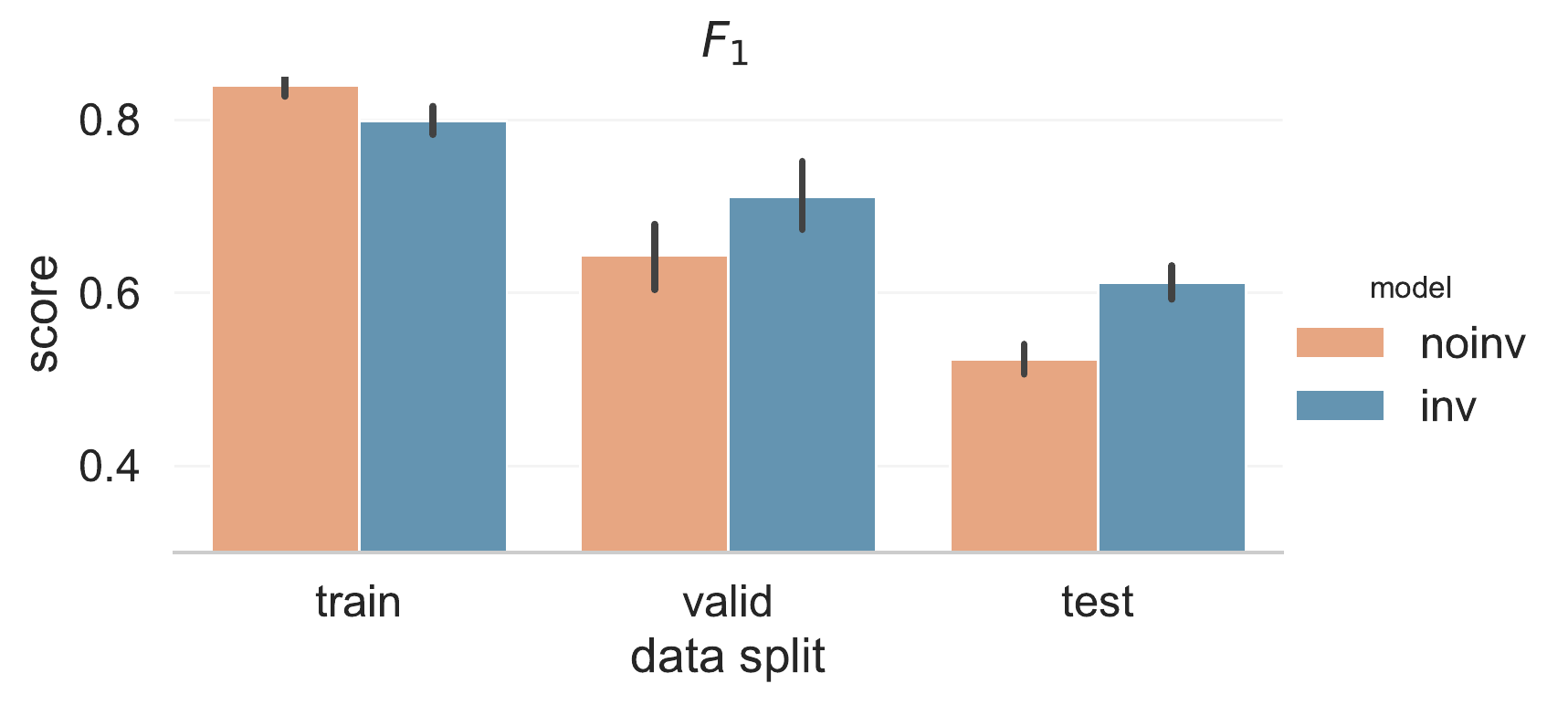}
    \caption{Results of the experiment on music data in terms of F-measure. Track scores are used to compute the average and the confidence intervals at 95\% (using bootstrapping). The proposed tempo-invariant architecture is able to better generalise over unseen data with respect to its standard CNN counterpart.}
    \label{fig:tempoinv_realdata}
\end{figure}

In this experiment we used real music recordings. We trained on an internal dataset (1368 excerpts from a variety of genres, summing up to 10 hours of music) and the RWC dataset \cite{goto2002rwc} (Popular, Genre and Jazz subsets) 
% and Beatles \cite{davies2009evaluation} datasets 
and tested on Ballroom \cite{gouyon2006experimental,krebs2013rhythmic} and GTZAN \cite{tzanetakis2002musical} datasets. With respect to the previous experiment we used the same input features, but larger networks\footnote{In terms of number of channels, layers and convolution kernel sizes, optimized on the validation set.} because of the higher amount of information contained in fully arranged recordings, with \texttt{inv} having $170$k trainable parameters and \texttt{noinv} $340$k.

The results in Fig.~\ref{fig:tempoinv_realdata} show that the proposed tempo-invariant architecture is performing worse on the training set, but better on the validation and test set, with the comparisons on train and test set being statistically significant ($p < 0.001$). Here the tempo-invariant architecture seems to act as a regularisation, allocating the network capacity to learning patterns that better generalise on unseen data, instead of fitting to the training set.

\section{Discussion}
\label{sec:discussion}

Since musicians are relentlessly creative, previously unseen rhythmic patterns keep being invented, much like ``out-of-vocabulary'' words in natural language processing \cite{schick2019learning}. As a result, the generalisation power of tempo-invariant approaches is likely to remain useful. Once tuned for optimal input representation and network capacity we expect tempo-invariant models to have an edge particularly on new, non-public test datasets.

Disentangling timbral pattern and tempo may also be useful to tasks such as auto-tagging: models can learn that some classes have a single precise tempo (e.g.\ ballroom dances \cite{gouyon2006experimental}), some have varying tempos within a range (e.g.\ broader genres or moods), and others still are completely invariant to tempo (e.g.\ instrumentation).

\section{Conclusions}
\label{sec:conclusions}

We introduced a scale-invariant convolution layer and used it as the main component of our tempo-invariant neural network architecture for downbeat tracking. We experimented on drum grooves and real music data, showing that the proposed architecture generalises to unseen tempi by design and achieves higher accuracy with lower capacity compared to a standard CNN.

% For bibtex users:
\bibliography{refs}

\begin{thebibliography}{10}
\providecommand{\url}[1]{#1}
\csname url@samestyle\endcsname
\providecommand{\newblock}{\relax}
\providecommand{\bibinfo}[2]{#2}
\providecommand{\BIBentrySTDinterwordspacing}{\spaceskip=0pt\relax}
\providecommand{\BIBentryALTinterwordstretchfactor}{4}
\providecommand{\BIBentryALTinterwordspacing}{\spaceskip=\fontdimen2\font plus
\BIBentryALTinterwordstretchfactor\fontdimen3\font minus
  \fontdimen4\font\relax}
\providecommand{\BIBforeignlanguage}[2]{{%
\expandafter\ifx\csname l@#1\endcsname\relax
\typeout{** WARNING: IEEEtran.bst: No hyphenation pattern has been}%
\typeout{** loaded for the language `#1'. Using the pattern for}%
\typeout{** the default language instead.}%
\else
\language=\csname l@#1\endcsname
\fi
#2}}
\providecommand{\BIBdecl}{\relax}
\BIBdecl

\bibitem{thaut2014human}
M.~Thaut, P.~Trimarchi, and L.~Parsons, ``Human brain basis of musical rhythm
  perception: common and distinct neural substrates for meter, tempo, and
  pattern,'' \emph{Brain sciences}, vol.~4, no.~2, pp. 428--452, 2014.

\bibitem{higgins2018towards}
I.~Higgins, D.~Amos, D.~Pfau, S.~Racaniere, L.~Matthey, D.~Rezende, and
  A.~Lerchner, ``Towards a definition of disentangled representations,''
  \emph{arXiv preprint arXiv:1812.02230}, 2018.

\bibitem{lecun1995convolutional}
Y.~LeCun, Y.~Bengio \emph{et~al.}, ``Convolutional networks for images, speech,
  and time series,'' \emph{The handbook of brain theory and neural networks},
  vol. 3361, no.~10, 1995.

\bibitem{bittner2017deep}
R.~M. Bittner, B.~McFee, J.~Salamon, P.~Li, and J.~P. Bello, ``Deep salience
  representations for {F0} estimation in polyphonic music,'' in \emph{Proc. of
  the International Society for Music Information Retrieval Conference
  (ISMIR)}, 2016, pp. 63--70.

\bibitem{schluter2018zero}
J.~Schl{\"u}ter and B.~Lehner, ``Zero-mean convolutions for level-invariant
  singing voice detection,'' in \emph{Proc. of the International Society for
  Music Information Retrieval Conference (ISMIR)}, 2018, pp. 321--326.

\bibitem{davies2005beat}
M.~E. Davies and M.~D. Plumbley, ``Beat tracking with a two state model [music
  applications],'' in \emph{Proc. of the International Conference on Acoustics,
  Speech and Signal Processing (ICASSP)}, vol.~3.\hskip 1em plus 0.5em minus
  0.4em\relax IEEE, 2005, pp. iii--241.

\bibitem{ellis2007beat}
D.~P. Ellis, ``Beat tracking by dynamic programming,'' \emph{Journal of New
  Music Research}, vol.~36, no.~1, pp. 51--60, 2007.

\bibitem{klapuri2005analysis}
A.~P. Klapuri, A.~J. Eronen, and J.~T. Astola, ``Analysis of the meter of
  acoustic musical signals,'' \emph{IEEE Transactions on Audio, Speech, and
  Language Processing}, vol.~14, no.~1, pp. 342--355, 2005.

\bibitem{degara2011reliability}
N.~Degara, E.~A. R{\'u}a, A.~Pena, S.~Torres-Guijarro, M.~E. Davies, and M.~D.
  Plumbley, ``Reliability-informed beat tracking of musical signals,''
  \emph{IEEE Transactions on Audio, Speech, and Language Processing}, vol.~20,
  no.~1, pp. 290--301, 2011.

\bibitem{giorgi2016multipath}
B.~Di~Giorgi, M.~Zanoni, S.~B{\"o}ck, and A.~Sarti, ``Multipath beat
  tracking,'' \emph{Journal of the Audio Engineering Society}, vol.~64, no.
  7/8, pp. 493--502, 2016.

\bibitem{papadopoulos2010joint}
H.~Papadopoulos and G.~Peeters, ``Joint estimation of chords and downbeats from
  an audio signal,'' \emph{IEEE Transactions on Audio, Speech, and Language
  Processing}, vol.~19, no.~1, pp. 138--152, 2010.

\bibitem{krebs2016downbeat}
F.~Krebs, S.~B{\"o}ck, M.~Dorfer, and G.~Widmer, ``Downbeat tracking using beat
  synchronous features with recurrent neural networks,'' in \emph{Proc. of the
  International Society for Music Information Retrieval Conference (ISMIR)},
  2016, pp. 129--135.

\bibitem{davies2006spectral}
M.~E. Davies and M.~D. Plumbley, ``A spectral difference approach to downbeat
  extraction in musical audio,'' in \emph{2006 14th European Signal Processing
  Conference}.\hskip 1em plus 0.5em minus 0.4em\relax IEEE, 2006, pp. 1--4.

\bibitem{durand2015downbeat}
S.~Durand, J.~P. Bello, B.~David, and G.~Richard, ``Downbeat tracking with
  multiple features and deep neural networks,'' in \emph{Proc. of the
  International Conference on Acoustics, Speech and Signal Processing
  (ICASSP)}.\hskip 1em plus 0.5em minus 0.4em\relax IEEE, 2015, pp. 409--413.

\bibitem{durand2016feature}
------, ``Feature adapted convolutional neural networks for downbeat
  tracking,'' in \emph{Proc. of the International Conference on Acoustics,
  Speech and Signal Processing (ICASSP)}.\hskip 1em plus 0.5em minus
  0.4em\relax IEEE, 2016, pp. 296--300.

\bibitem{goto2001audio}
M.~Goto, ``An audio-based real-time beat tracking system for music with or
  without drum-sounds,'' \emph{Journal of New Music Research}, vol.~30, no.~2,
  pp. 159--171, 2001.

\bibitem{dixon2001automatic}
S.~Dixon, ``Automatic extraction of tempo and beat from expressive
  performances,'' \emph{Journal of New Music Research}, vol.~30, no.~1, pp.
  39--58, 2001.

\bibitem{eck2007beat}
D.~Eck, ``Beat tracking using an autocorrelation phase matrix,'' in \emph{Proc.
  of the International Conference on Acoustics, Speech and Signal Processing
  (ICASSP)}, vol.~4.\hskip 1em plus 0.5em minus 0.4em\relax IEEE, 2007, pp.
  IV--1313.

\bibitem{goto2011songle}
M.~Goto, K.~Yoshii, H.~Fujihara, M.~Mauch, and T.~Nakano, ``Songle: A web
  service for active music listening improved by user contributions,'' in
  \emph{Proc. of the International Society for Music Information Retrieval
  Conference (ISMIR)}, 2011, pp. 311--316.

\bibitem{krebs2015inferring}
F.~Krebs, A.~Holzapfel, A.~T. Cemgil, and G.~Widmer, ``Inferring metrical
  structure in music using particle filters,'' \emph{IEEE/ACM Transactions on
  Audio, Speech, and Language Processing}, vol.~23, no.~5, pp. 817--827, 2015.

\bibitem{bock2011enhanced}
S.~B{\"o}ck and M.~Schedl, ``Enhanced beat tracking with context-aware neural
  networks,'' in \emph{Proc. of the International Conference on Digital Audio
  Effects (DAFx)}, 2011, pp. 135--139.

\bibitem{bock2016joint}
S.~B{\"o}ck, F.~Krebs, and G.~Widmer, ``Joint beat and downbeat tracking with
  recurrent neural networks,'' in \emph{Proc. of the International Society for
  Music Information Retrieval Conference (ISMIR)}, 2016, pp. 255--261.

\bibitem{korzeniowski2014probabilistic}
F.~Korzeniowski, S.~B{\"o}ck, and G.~Widmer, ``Probabilistic extraction of beat
  positions from a beat activation function,'' in \emph{Proc. of the
  International Society for Music Information Retrieval Conference (ISMIR)},
  2014, pp. 513--518.

\bibitem{xu2014scale}
Y.~Xu, T.~Xiao, J.~Zhang, K.~Yang, and Z.~Zhang, ``Scale-invariant
  convolutional neural networks,'' \emph{arXiv preprint arXiv:1411.6369}, 2014.

\bibitem{kanazawa2014locally}
A.~Kanazawa, A.~Sharma, and D.~Jacobs, ``Locally scale-invariant convolutional
  neural networks,'' in \emph{Deep Learning and Representation Learning
  Workshop, Neural Information Processing Systems (NIPS)}, 2014.

\bibitem{krebs2015efficient}
F.~Krebs, S.~B{\"o}ck, and G.~Widmer, ``An efficient state-space model for
  joint tempo and meter tracking,'' in \emph{Proc. of the International Society
  for Music Information Retrieval Conference (ISMIR)}, 2015, pp. 72--78.

\bibitem{groove2019}
J.~Gillick, A.~Roberts, J.~Engel, D.~Eck, and D.~Bamman, ``Learning to groove
  with inverse sequence transformations,'' in \emph{Proc. of the International
  Conference on Machine Learning (ICML)}, 2019.

\bibitem{davies2009evaluation}
M.~E. Davies, N.~Degara, and M.~D. Plumbley, ``Evaluation methods for musical
  audio beat tracking algorithms,'' \emph{Queen Mary University of London,
  Centre for Digital Music, Tech. Rep. C4DM-TR-09-06}, 2009.

\bibitem{goto2002rwc}
M.~Goto, H.~Hashiguchi, T.~Nishimura, and R.~Oka, ``{RWC} music database:
  Popular, classical and jazz music databases,'' in \emph{Proc. of the
  International Society for Music Information Retrieval Conference (ISMIR)},
  vol.~2, 2002, pp. 287--288.

\bibitem{gouyon2006experimental}
F.~Gouyon, A.~Klapuri, S.~Dixon, M.~Alonso, G.~Tzanetakis, C.~Uhle, and
  P.~Cano, ``An experimental comparison of audio tempo induction algorithms,''
  \emph{IEEE Transactions on Audio, Speech, and Language Processing}, vol.~14,
  no.~5, pp. 1832--1844, 2006.

\bibitem{krebs2013rhythmic}
F.~Krebs, S.~B{\"o}ck, and G.~Widmer, ``Rhythmic pattern modeling for beat and
  downbeat tracking in musical audio,'' in \emph{Proc. of the International
  Society for Music Information Retrieval Conference (ISMIR)}, 2013, pp.
  227--232.

\bibitem{tzanetakis2002musical}
G.~Tzanetakis and P.~Cook, ``Musical genre classification of audio signals,''
  \emph{IEEE Transactions on Speech and Audio Processing}, vol.~10, no.~5, pp.
  293--302, 2002.

\bibitem{schick2019learning}
T.~Schick and H.~Sch{\"u}tze, ``Learning semantic representations for novel
  words: Leveraging both form and context,'' in \emph{Proceedings of the AAAI
  Conference on Artificial Intelligence}, vol.~33, 2019, pp. 6965--6973.

\end{thebibliography}

% For non bibtex users:
%\begin{thebibliography}{citations}
%
%\bibitem {Author:00}
%E. Author.
%``The Title of the Conference Paper,''
%{\it Proceedings of the International Symposium
%on Music Information Retrieval}, pp.~000--111, 2000.
%
%\bibitem{Someone:10}
%A. Someone, B. Someone, and C. Someone.
%``The Title of the Journal Paper,''
%{\it Journal of New Music Research},
%Vol.~A, No.~B, pp.~111--222, 2010.
%
%\bibitem{Someone:04} X. Someone and Y. Someone. {\it Title of the Book},
%    Editorial Acme, Porto, 2012.
%
%\end{thebibliography}

\end{document}